# FIT: Tag based method for fusion proteins identification

Kang Ning[1,3], Alexey I. Nesvizhskii [1,2,*]

[1] Department of Pathology, University of Michigan, Ann Arbor, MI 48109
[2] Center for Computational Biology and Medicine, University of Michigan, Ann Arbor, MI 48109
[3] Qingdao Institute of Bioenergy and Bioprocess Technology, Chinese Academy of Sciences, Qingdao, Shandong, China, 266101

AUTHOR EMAIL ADDRESS:   nesvi@med.umich.edu

TITLE RUNNING HEAD：   Tag based method for Fusion proteins Identification

CORRESPONDING AUTHOR FOOTNOTE：
Alexey I. Nesvizhskii
Department of Pathology
University of Michigan Medical School
1301 Catherine
4237 Medical Science I
Ann Arbor, MI 48109

**Phone:** 734.764.3516
**Email:** Alexey@umich.edu

**ABSTRACT**
There is increased interest in the identification and analysis of gene fusions and chimeric RNA transcripts. While most recent efforts focused on the analysis of genomic and transcriptomic data, identification of novel peptides corresponding to such events in mass spectrometry-based proteomic datasets would provide complementary, protein-level evidence. The process of identifying fusion proteins from mass spectrometry data is inherently difficult because such events are rare. It is also complicated due to large amount of spectra collected and the explosion in the number of candidate peptide sequences that need to be considered, which makes exhaustive search for all possible fusion partner proteins impractical. In this work, we present a sequence tag based fusion protein identification algorithm, FIT, that combines the virtue of both *de novo* sequence tag retrieval and peptide-spectrum matching for identification of fusion proteins. Results on simulated datasets show high sensitivity and low false positive rates for fusion protein identification by the FIT algorithm.

## 1 INTRODUCTION

Fusion genes are formed through the rearrangement of two or more genes, which originally coded for separate proteins. Since fusion genes are often correlated with tumor biomarkers in cancer gene architecture [1,2], identification of fusion genes significantly contributes to the understanding of cancer progression. Most efforts naturally focused on the identification of gene fusions and chimeric transcripts using genomics data [3,4,5-7]. In particular, multiple bioinformatics tools for detection of chimeric transcripts from RNA-Seq data have been recently described [8-11].

Identification of fusion (or chimeric) proteins, i.e. protein-level products resulting from gene fusions and other types of chimeric transcripts from tandem mass spectrometry (MS/MS) data could serve as an alternative or supplementary method for detection of such events [12]. A fusion protein is represented by $P_{i-p}||P_{j-s}$, in which $P_{i-p}$ and $P_{j-s}$ are prefix and suffix of protein $P_i$ and $P_j$, respectively. Protein $P_i$ and $P_j$ are the protein products of two corresponding genes, and they are referred to as the "fusion partner proteins". Each pair of fusion partner proteins corresponds to a fusion event, and the position in fusion protein where the two proteins "fuse" is referred to as the protein fusion site (breakpoint). The peptide that straddles the fusion site is referred to as a fusion (or chimeric) peptide. From proteomic experiment point of view, MS/MS datasets acquired on biological samples containing fusion proteins may contain spectra representing fusion peptides. Therefore, computational analysis of MS/MS datasets could lead to the identification of fusion peptides and their corresponding fusion proteins.

Many fusion proteins are already known, and more could be identified with confidence from e.g. RNA-Seq data when both RNA-Seq and MS/MS data are available for the same biological system. These known or predicted fusion peptide sequences can be added to the protein sequences database, and MS/MS spectra can then be searched against this protein database in a convention way[13]. However, the identification of truly novel fusion peptides from MS/MS data alone is difficult and so far has been attempted in only several works. The first, targeted approach [12] is based on *de novo* algorithm to identify peptide sequences, followed by alignment of these peptides against a limited set of protein sequences. In this work, the abundance of the fusion protein in the sample was increased via immunoprecipitation (IP) with an antibody against one of the fusion protein partners, followed by MS analysis (IP-MS). Furthermore, multiple proteases were used to digest proteins to generate peptides overlapping the same fusion site, thus increasing the confidence in the identification. The computational method used in this work assumed that at least one of the fusion partners is known *a priori*, and it cannot be effectively extended to less restrictive situations. The second approach is global in that it searches for all possible fusion sites. In this method, based on spectrum alignment [14,15], fragments of protein sequences are transformed into theoretical spectra. The experimental MS/MS spectra are aligned against theoretical spectra of every possible combination of fragments in the protein database. This method essentially performs an unrestricted search of all possible candidate fusion peptides (i.e. without restricting the analysis to any specific subset of proteins or spectra). As such it is computationally expensive and also may result in a high rate of false positive fusion peptide identifications.

In this paper, we proposed a new algorithm, FIT, for sequence tag based fusion protein identification. FIT is designed to provide a solution that bridges the targeted and global approaches described above. It implements a protein sequence filtration step that restricts the fusion peptide search space by requiring that at least one of the candidate fusion partners is identified in the sample with high confidence by known (i.e. non-fusion) peptides. This assumption reflects our expectation that fusion peptide identification would be most successful in the case of experiments that enrich the abundance of otherwise low abundant fusion proteins (e.g. in IP-MS experiments in which the fusion protein is the bait protein or one of the interacting partners of the bait). Furthermore, MS/MS spectra are also filtered to keep only spectra that cannot be assigned with confidence to known protein sequences. Another key feature of FIT algorithm is the selection of candidate fusion partner proteins using sequence tags extracted from the spectra [16-20]. In FIT, a set of tags are generated for each of the unassigned spectra, and then aligned to candidate proteins. A pair of candidate fusion partner proteins is selected only when a pair of tags from a single spectrum maps to these two proteins. Then FIT scores candidate fusion peptides via full spectrum-peptide matching.

The advantage of the proposed method is that fusion protein identifications are based on only a small subset of proteins and a small subset of all spectra. Using sequence tags for candidate fusion peptide retrieval reduces the computational time and increases the sensitivity of peptide identification. Sequence tags generated from spectra are more reliable than *de novo* sequencing results, and the alignment of tags to protein sequences is followed up by conventional spectrum-peptide matching. Therefore, FIT combines the virtue of both *de novo* tag retrieval and peptide-spectrum matching.

## 2 METHODS

### 2.1 Experiment data

The primary goal of this work is to introduce the new method and to evaluate its performance. The evaluation of the method with respect to its ability to find novel fusion peptides in biological context is not feasible without substantial work to experimentally validate novel predictions. Thus, the performance of the algorithm is tested here using simulated experiments based on real MS/MS spectra and simulated fusion events. A similar strategy was used in[12].

MS/MS spectra were taken from a publicly available Human T Leukemic Cells dataset [21], the first replicate of Whole Cell Lysate (WCL) fraction experiment. The spectral dataset used here contained a total of 269,371 MS/MS spectra. The spectra were searched by X! TANDEM [22] (with k-score [23]) against the SwissProt database [24] (Release 57.10, appended with reversed protein sequences of equal size as decoys), or against the modified database (referred to as "fusion database" below) created as described below. The search parameters in all searches were: 2.0 Da parent mass tolerance, 0.8 Da fragment monoisotopic mass tolerance, and allowing tryptic peptide only. Two modifications were considered as variable modifications: Met oxidation and N-terminal acetylation. Peptide to spectrum matches were analyzed using PeptideProphet and ProteinProphet [25], which posterior probabilities to peptide and protein identifications, respectively (described in more detail below).

## 2.2 Simulated fusion database construction

To create the simulated fusion database, we first selected top 200 proteins with highest abundances and with ProteinProphet probabilities (> 0.9) based on the initial search of the experimental MS/MS data against the regular SwissProt database (see Experimental Data). These proteins were considered as "fusion proteins" (ground truth) to be identified by searching MS/MS spectra against this simulated fusion protein database. Second, 200 unique peptides, each being selected from one of the 200 proteins, were extracted, and preference was given to peptides located near the middle of the protein sequences. These peptides represent the "fusion peptides" to be identified. Third, each of the 200 fusion proteins was broken at a breakpoint (referred to as "fusion sites"). The fusion site was set either at the middle of the corresponding fusion peptide, or shifted with respect to the middle of the peptide by one or more amino acids in either direction. The fusion protein was then translocated with a fragment of another protein randomly selected from the list of all identified proteins (Figure 1). This procedure reflects in part the real situation in which many genes involved in fusion recombine with different partners [26]. Each pair of proteins thus constructed is referred to as a pair of "fusion partner proteins". Then the proteins that are used to construct these fusion partner proteins are removed from the database, and the new fusion partner proteins are inserted (Figure 1).

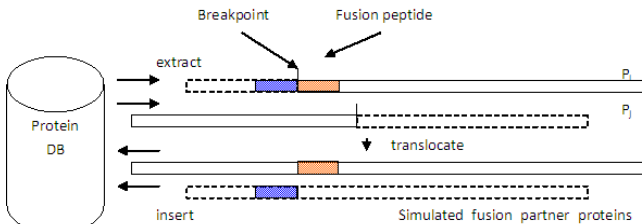

**Figure 1. An illustration of creating one pair of fusion partner proteins based on proteins $P_i$ and $P_j$.** Shaded fractions represent the fusion peptide for this fusion event. Both $P_i$ (one of the top 200 proteins based on abundance) and $P_j$ (randomly selected among all identified proteins) are extracted from protein database, translocated, and the resulting two proteins are inserted back in the protein database.

## 2.3 The FIT algorithm

FIT algorithm implements the following major steps: filtering of protein sequences and MS/MS spectra; selection of fusion partner proteins by sequence tagging; scoring of MS/MS spectra against candidate fusion peptides.

*2.2.1 Phase I, Restrict the search space for fusion partner proteins and protein fusion site* In the initial database search, all spectra (referred to as {S}) are searched by X! TANDEM [22] against protein sequences database (referred to as {P}) for peptide identification. Then PeptideProphet [27] (a tool to automatically validate peptide assignments to MS/MS spectra made by database search, a part of the TPP suite [28]) is used to assign a probability to each of the peptide to spectrum matches. ProteinProphet [25] then performs protein level inference and computes a protein probability.

Selected peptides (with a high PeptideProphet probability) are mapped to protein sequences, and the protein coverage is computed as the proportion of the sequence covered by the identified peptides. The set of proteins {$P_{selected}$} selected as fusion protein candidates can be represented as:

$$\{P_{selected}\} = \{P \mid ProteinProphet(P) > 0.9 \ \& \ coverage(P) > 0.5\} \quad (1)$$

To restrict the number of proteins to be searched for computational efficiency, and more importantly to ensure reliable (low error rate) identification of fusion peptides, it is required that at least one of the two candidate fusion partner proteins is in the selected set described above. Additionally, it is assumed that there is no "known" peptide identified in the dataset from either one of the fusion partner proteins that overlaps with the fusion site. Furthermore, only fragments that do not have peptide alignments could contain possible protein fusion site (in general, these assumptions may not always be true; the cells can express both protein forms, i.e. the canonical protein and the chimera form).

*2.2.2 Phase II, Restricting the number of candidate spectra*
To restrict the number of candidate spectra that may identify fusion peptides, spectra that can be assigned to peptides from known proteins with high confidence (with a probability threshold corresponding to false discovery rate, FDR, below 0.05) are filtered out. Then, a spectrum quality score (SQS) [29], which is a score that takes into consideration multiple features reflecting MS/MS spectrum quality (total number of peaks, signal to noise level, etc.), is computed for each spectrum. Spectra with SQS scores above 1.0 are considered high quality spectra[29]. Those spectra that have low PeptideProphet probability (below the threshold) and SQS score above 1.0 are considered unassigned high quality spectra. A fusion peptide is more likely to be identified from this selected set of spectra, denoted as {$S_{selected}$}:

$$\{S_{selected}\} = \{S \mid PeptideProphet(S) < threshold \ \& \ SQS(S) > 1.0\} \quad (2)$$

*2.2.3 Phase III, Selection of candidate fusion proteins by tag*
Since a fusion peptide straddles the protein fusion site, it should have its prefix matching one protein, and its suffix matching another protein. Two proteins, $P_i$ and $P_j$ are selected as putative fusion partner proteins when there is a spectrum S (from the selected set, see E. 2) from which we can generate two tags $T_i$ and $T_j$, with $T_i$ (and its prefix mass) matching $P_i$ and $T_j$ (and its suffix mass) matching $P_j$. An illustration of this tag mapping process is shown in Figure 2. The mapping of sequence tags as well as prefix and suffix masses are based on the PepLine [30] software, which could map peptide sequence tags (PST, consisting of tag sequence and its prefix or suffix masses) from the spectrum onto protein sequences.

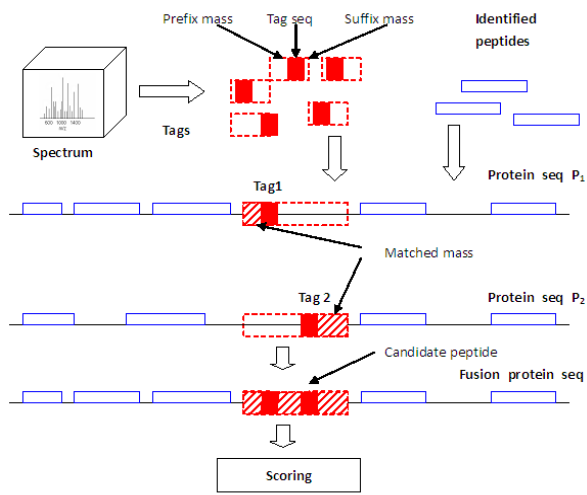

**Figure 2. An illustration of putative fusion partner protein selection based on extracted sequence tags.** Solid boxes indicate assigned peptides, Thick bars indicate tag sequences and dashed boxes indicate tags that include prefix and suffix masses.

*2.2.4   Phase IV, Identification of fusion protein* The candidate fusion peptides nominated as described above are scored via direct spectrum-peptide comparison. The dot product score is computed using X! Tandem for each spectrum S and its corresponding candidate fusion peptide $P_{i-p}||P_{j-s}$. Good matches are expected to have a large dot product  (generally > 300); as a set, the list of accepted fusion peptides should have a small FDR (e.g. < 0.01) [31]. At this stage, experiment specific properties such as the specificity of the proteolytic enzyme used to digest proteins can be used to eliminate candidate peptides containing missed cleavages or not conforming to the enzymatic cleavage rules at the N- or C-terminus. Additionally, fusion peptides could be required to be identified from multiple MS/MS spectra.

## 3   RESULTS AND DISCUSSIONS

The primary goal of this work is to introduce the new method and to evaluate its performance. The evaluation of the method with respect to its ability to find novel fusion peptides in biological context is not feasible without substantial work to experimentally validate novel predictions. Thus, the performance of the algorithm is tested here using simulated experiments based on real MS/MS spectra and simulated fusion events. A similar strategy was used in[12].

### 3.1   Experiment data and simulated fusion database

MS/MS spectra were taken from a publicly available Human T Leukemic Cells dataset [21] (see Methods for detail). The spectral dataset, containing 269,371 MS/MS spectra, was first searched by X! TANDEM [22] (with k-score [23]) against the SwissProt database [24] (Release 57.10, appended with reversed protein sequences of equal size as decoys).  Peptide to spectrum matches were analyzed using PeptideProphet and ProteinProphet [25], leading to the identification of 1,936 SwissProt protein entries with high ProteinProphet probabilities (decoy-estimated FDR of ~ 0.01). Among the 224,614 spectra that were initially unassigned, 7,098 spectra are of high quality (SQS > 1.0).

The simulated fusion protein database was created as follows (see Method for details). The protein abundance of each of the identified protein was estimated using spectrum counts [32]. Proteins were then sorted by their abundance, and the top 200 proteins with highest abundances were selected as fusion proteins. Then, 200 unique "fusion peptides" were selected from these proteins, corresponding to a total of 521 spectra (referred to as "positive spectra dataset") – a small fraction compared to the total of 269,371 spectra in this dataset. Each of these "fusion proteins" were broken at a breakpoint (set at the middle of the corresponding fusion peptide), and translocated with another protein (not necessarily identified with high probability). This process resulted in 180 randomly selected proteins to be translocated with 200 selected high abundance proteins to construct "fusion partner proteins". The original proteins that were used to construct these "fusion partner proteins" were removed from the protein database, and the "fusion partner proteins" were inserted in it instead. The fusion protein database thus constructed contained 20,402 proteins, in which 20,042 proteins are identical to the ones in the original protein database. By this construction, positive spectra do not have the corresponding peptides in simulated fusion protein database, but they could be identified as fusion peptides by FIT method.

It is acknowledged that the simulated fusion database generation procedure described above (and also in the FIT algorithm) makes a number of simplifying assumptions regarding the fusion protein identification problem. Nevertheless, the key assumptions are not without the merit. Most importantly, it is assumed that at least one of the proteins involved in fusion is detected with high sequence coverage. As discussed in the Introduction, we believe that it is most realistic to search for fusion peptides in datasets generated in experiments in which fusion proteins are enriched, such as IP-MS experiments targeting a protein (and its interacting partners) from a particular (often cancer-related) pathway of interest. The computational tests performed also reflect another major difficulty, the fact that only a tiny fraction of all acquired MS/MS spectra in the dataset correspond to fusion peptides.

### 3.2   Candidate fusion protein selection and identification

*3.2.1   Phase I, Filtration of proteins*
All spectra from WCL dataset were re-searched against the simulated fusion protein database in the same way as in the original search. Based on this database search, 2,458 protein entries were identified with high ProteinProphet probabilities (> 0.9; corresponding FDR of 0.009). Additionally, it is noticed that out of the originally identifiable 1,936 proteins, 1,754 (90.6%) untranslocated proteins would still be identified from searching the simulated fusion protein database (Table 1). This indicates that the construction of the simulated fusion protein database does not significantly change the set of proteins that could be identified with high ProteinProphet probability.

We first analyzed the effect of protein filtration on fusion protein selection. There are two measurements for this analysis: the proportion of fusion partner proteins that can generate correct fusion partner protein pairs (% of fusion partner proteins) after filtration, and the number of fusion events that could be identified after filtration (number of fusion proteins). We have first analyzed the effect of applying different ProteinProphet probability threshold on one or both of the fusion partner proteins. Applying ProteinProphet probability threshold of 0.9 on one of the fusion partner proteins, 95% of fusion partner proteins that pass the filtration could form fusion partner protein pairs, and 90% (180/200) of fusion proteins could be still identified after filtration. When this ProteinProphet probability threshold was applied to both of the fusion partner proteins, there were 176 and 173 fusion partner proteins that contained the prefix and suffix of correct fusion proteins and passed the filtration; 84% (167/200) of fusion proteins could be identified from these fusion partner proteins (Table 1). In other words, in this dataset requiring that both of the fusion partner proteins pass the probability threshold of 0.9 did not significantly reduce the number of identifiable fusion protein, whereas the number of possible combinations of fusion partner proteins was reduced drastically.

Another protein filtration step that we implemented is based on requiring certain minimum protein sequence coverage by identified peptides. Based on fusion partner proteins that passed the ProteinProphet probability threshold (0.9), it was observed that as the required protein coverage increased, the number of true fusion proteins that could be identified was consistently decreasing (Figure 3). At the same time, the total number of candidate fusion partner pairs was decreasing as well. As a tradeoff, we selected the protein coverage threshold of 0.5 (required for at least one of the two fusion partner candidates) as a filter. It should be noted that the protein filters described above are likely to be dataset specific, and may have to be adjusted for different dataset.

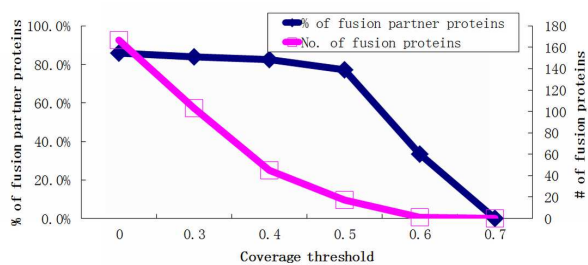

**Figure 3. The effect of protein coverage threshold on the filtration of fusion proteins.** Results correspond to requiring that at least one of the two fusion partner proteins pass a specific protein coverage threshold (in addition to the ProteinProphet probability threshold of 0.9 applied on both fusion partner proteins). "% of fusion partner proteins" is computed by dividing the number of fusion partner proteins after filtration that can generate correct fusion partner protein pairs by the total number of fusion partner proteins after filtration.

### 3.2.2 Phase II, Filtration of spectra

Apart from protein filtration, MS/MS spectra are also filtered. Based on PeptideProphet probability ($> 0.7$ are considered as assigned; FDR = 0.03) and spectrum quality score (SQS $> 1.0$ as high quality) assigned to each of the spectra, there were 231,608 unassigned spectra, of which 8,712 were of high quality. These numbers are quite similar to the results of searching the original protein database. From the results of this database search, it was also observed that out of the 521 spectra in positive dataset (ground truth), 6 are charge 1 spectra, and 20 were assigned to peptides in the fusion protein database (i.e., there was another, likely homologous, protein in the database containing the same sequence). The remaining 495 spectra were unassigned spectra, corresponding to 178 fusion proteins, out of which 167 fusion proteins had both fusion partner proteins identified with ProteinProphet probability $> 0.9$ (Table 1). 391 out of these 495 spectra were of high quality, corresponding to 165 fusion proteins after protein filtration.

### 3.2.3 Phase III, Selecting candidate fusion proteins by tags

As the next step, we have analyzed the effect of fusion protein selection from fusion partner proteins using sequence tags. Previous studies have shown that first, for the majority of spectra, among top 100 tags, there are at least 2 length-2 tags that are correct; and among top 200 tags, there are at least 2 length-3 tags that are correct. Second, most of the tags are distributed near the center of the peptide sequences [33]. Therefore, in this study, we generated top 100 length-2 and top 200 length-3 tags for each of the spectra. When mapping tags on fusion partner proteins, it is discovered that based on length-3 tags, there are only 163 spectra (out of 495 spectra) from positive dataset for which tags can be mapped to both fusion partner proteins. Based on length-2 tags, there are 227 spectra (out of 495 spectra) from positive datasets for which tags can be mapped to both prefix and suffix of the fusion proteins (Table 1). Thus, using length-2 tags allows selection of a higher percentage of true fusion peptides than length-3 tags.

The tag-based selection of fusion protein candidates resulted in approximately 30% loss in the number of possible fusion protein identifications (in contrast, the protein filtration step resulted in the loss of 37 fusion proteins, or less than 20%), see Table 1. We thus further investigated the reason for the 30% loss at the sequence tag-based filtering stage. We found that when the fusion breakpoint is in the middle of the fusion peptide sequence (as we implemented as a part of our procedure for creating the simulated fusion protein database), many of the extracted sequence tags cannot be mapped to either prefix or suffix of the fusion peptide (i.e., there is no pair of tags that could map to the prefix and suffix of the corresponding protein fusion candidate sequences).

To further assess the effect of breakpoint position on candidate fusion protein selection, the analysis was repeated using different sets of fusion partner proteins with breakpoints shifted (either left or right) with respect to the middle of the fusion peptide sequence (keeping others settings the same). Results based on different breakpoints (defined as the offset from the middle of the peptide) of the fusion proteins are shown in Figure 4. It is interesting to note that offset of 1 amino acid would yield largest number of fusion event detection. Specifically, based on length-3 tags, 294 spectra in positive dataset can be assigned to correct fusion proteins if the fusion partner proteins are created with breakpoint offset middle of peptide by 1 amino acid. Based on length-2 tags, 471 spectra in positive dataset can be assigned to correct fusion proteins if the fusion partner proteins are created with breakpoint offset middle of peptide by 1 amino acid. The main reason for significant increase of the number of spectra that could be retrieved by tags is that the middle part of the peptide is usually fragmented more completely than other parts of the peptide [33], and therefore most of tags from spectra correspond to the middle part of the peptide. If breakpoint is set at the middle, then some of the tags are unmappable on either prefix or suffix of fusion peptide. Shifting $\geq 1$ amino acids from the middle could attribute these tags to either prefix or suffix of the fusion peptide, and therefore increase the number of spectra that could be selected by tags. Based on these spectra and 167 fusion proteins that could be identified after protein filtration, shifting breakpoint would yield sensitivity (#true positive / (#true positive + #false negative)) of 73.7% (123/167) based on length-3 tags, and sensitivity of 93.4% (156/167) based on length-2 tags for fusion protein selection. The overall sensitivity of 61.5% (123/200) and 78.0% (156/200) could be achieved based on length-3 tags and length-2 tags by the FIT algorithm, respectively (Table 1).

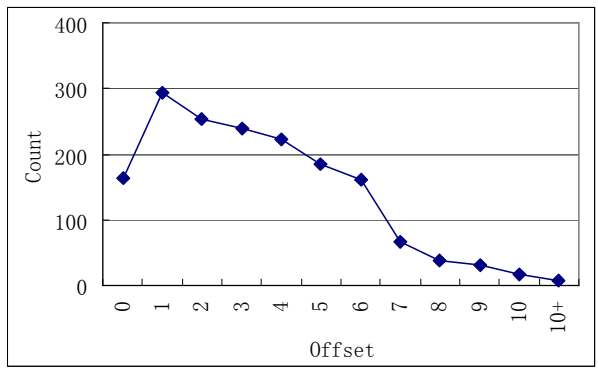

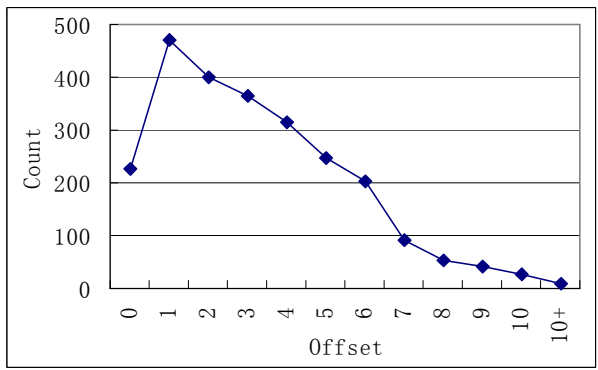

**Figure 4. The number of spectra in positive dataset that could be selected by tags, categorized by the offset of breakpoint from middle of the fusion peptide.** Shift values indicate the absolute shift values from the middle. Results are based on filtered fusion partner proteins for (a) length-3 tags and (b) length-2 tags.

**Table 1.** The number of spectra and proteins after each filtration and scoring step. After protein filtering, only results based on filtering both of the fusion partner proteins are used, and in peptide-spectrum scoring, only results based on length-2 tags are shown. HQ: Results based on using high quality (SQS>1) spectra only.

|  | Original | After protein filtering | After spectra filtering (FDR < 0.05) | After fusion partner protein matching by tags | After peptide-spectrum-match scoring (FDR < 0.01) |
|---|---|---|---|---|---|
| # of all spectra | 269,371 | - | 231,608 (HQ: 8,712) | Length-2 tag: 516 Length-3 tag: 451 | - |
| # of all proteins | 20,402 | 2,458 | - | - | - |
| # of spectra for fusion proteins | 521 | - | 495 (HQ: 391) | Length-2 tag: 471 (Breakpoint at middle: 227) Length-3 tag: 294 (Breakpoint at middle: 163) | Length-2 tag: 295 (HQ: 370) |
| # of fusion proteins identifiable | 200 | 167 (Filtering one partner: 180) | 167 (HQ: 165) | Length-2 tag: 156 Length-3 tag: 123 | Length-2 tag: 125 (HQ: 146) |

### 3.2.4 Phase IV, Scoring of fusion proteins

After fusion partner protein matching using extracted sequence tags, the number of candidate fusion proteins and candidate spectra to be considered has been drastically reduced. However, there are still a number of candidate fusion peptides presented (354 in total), of which a substantial proportion (198/354, or 56.0%) are false positives. Therefore, fusion peptides are further scored via the cross correlation analysis of the full experimental and theoretical peptide spectra. We have performed this scoring by searching $\{S_{selected}\}$ using X! Tandem (enzymatic cleavage rules considered in the search) against a protein database constructed as follows. We have collected all candidate fusion peptides (354 candidate peptides in total, of which 156 correspond to ground truth), and appended this collection of peptides to the list of 2,458 proteins identified with ProteinProphet probabilities > 0.9 in the earlier X! Tandem search against the simulated fusion protein database.

Two spectral sets were searched against this new database. One spectral set contains all initially unassigned spectra (a superset of $\{S_{selected}\}$) that are selected by length-2 tags (including 471 spectra in positive dataset), and another spectra dataset contains high quality unassigned spectra ($\{S_{selected}\}$) that are selected by length-2 tags (including 375 spectra in positive dataset). All matches between spectra in the positive dataset and their corresponding peptides were considered correct. Incorrect matches included matches of spectra in the positive dataset to peptides that are not their respective fusion peptides, as well as matches of spectra not in the positive datasets to fusion peptides. The peptide-spectrum matches were sorted by their X! Tandem dot product scores, and the FDR was computed based on the number of correct and incorrect matches passing a particular score threshold. Results show that the majority of fusion peptides could be identified with very small FDR (Figure 5). Based on high quality unassigned spectra, 146 true fusion peptides (out of 156) could be identified at FDR < 0.01. On the other hand, based on all initially unassigned spectra, only 125 fusion peptides could be identified at the same FDR (Table 1). Thus, searching high quality unassigned spectra (as compared to all spectra) improve the accuracy of fusion peptide identification.

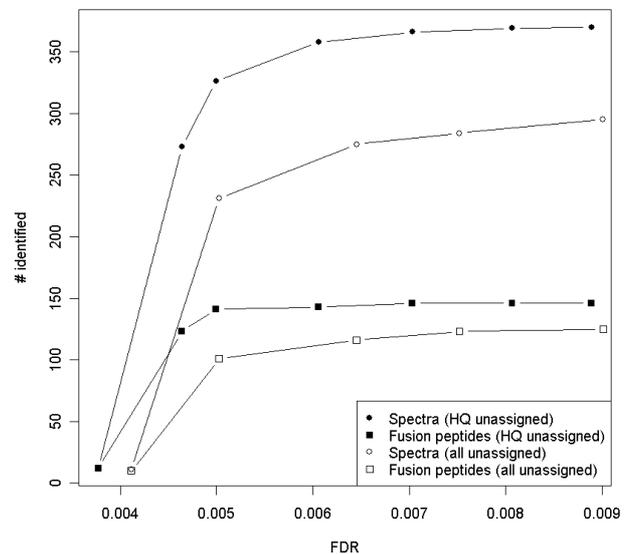

**Figure 5. The correlation of the number of correct fusion event identified and the FDR for fusion event identification.** The fusion event identifications are performed based on two spectra datasets: one containing all unassigned spectra, another containing high quality unassigned spectra.

We have also analyzed the relationship of the number of spectra with the likelihood of the fusion peptide identification. In the positive dataset, there were cases of multiple MS/MS spectra correspond to the same peptide (521 spectra for 200 unique fusion peptides). We have computed the number of spectra per unique fusion peptide, and analyzed its relationship with the identification rate of the fusion event. The identification ratio is defined here as the number of fusion peptide identified by FIT, over the total number of selected fusion peptides. Figure 6 shows that the identification ratio increases with the increasing number of spectra per peptide. For example, for fusion peptide identification based on length-3 tags, with the number of spectra per peptide equal to 1, the identification ratio is around 40%. When the number of spectra per peptide is greater than 1, the identification ratio is higher than 60%. This shows that, as expected, the sensitivity of fusion event identification is positively correlated with the number of spectra per peptide.

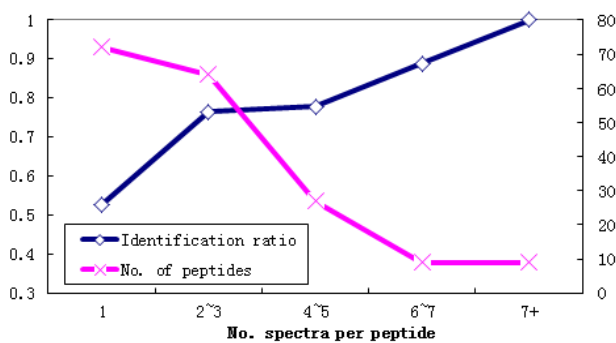

**Figure 6. The identification ratio of fusion events against the number of spectra per fusion peptide.** X-axis represents the number of spectra per peptide threshold for fusion peptide, and y-axis represents identification ratio. Results are based on filtered fusion partner proteins and length-2 tags.

## 4   CONCLUSIONS

Fusion protein identification from MS/MS data is a potentially valuable strategy for detection of such events at the protein level. However, current large-scale proteomic experiments generate large amounts of MS/MS spectra, making exhaustive search for fusion protein identification difficult. In this work, a sequence tag-based fusion protein identification algorithm, FIT, has been proposed that can select fusion partner proteins for more effective fusion peptide identification. Filtration of candidate proteins and candidate spectra are two key steps in the FIT algorithm, which significantly reduces the number of candidate spectra and candidate fusion partner proteins. To test the method, we relied on real MS/MS spectra searched against sequence databases containing simulated protein fusion events. In these tests, a relatively high sensitivity of fusion protein identification was achieved with low false discovery rate.

FIT algorithm (and the entire concept of searching for fusion events using proteomic data alone) has several limitations. First, FIT has been designed for a particular type of experiments, such as IP-MS protein interaction profiling experiments, in which the fusion proteins are expected to be enriched compared to their endogenous levels in the cell. For experiments in which the abundance level of fusion proteins remains low, filtration of protein sequences should be less stringent, and significantly more computational resources would then be needed for fusion partner protein selection and identification by FIT algorithm. Additionally, the analysis of fusion events has been far more successful at the genomic level. Databases of known fusion events already exists based on bioinformatics analysis of mRNA and EST sequences in the GenBank, manual collection of literature data, and integration of other known database [34]. Furthermore, RNA-seq datasets are now being generated with increasing frequency, allowing more comprehensive analysis of gene fusions and chimera transcripts that previously possible. Thus, the search for protein-level evidence of gene fusion events would need to be more tightly coupled with the analysis of matching transcriptome data generated in parallel with proteomic data using matching biological samples.

## ACKNOWLEDGEMENTS

This work was supported in part by NIH grants R01-CA-126239 and R01-GM-094231. And it was also supported in part by National Natural Science Foundation of China (NSFC Grant No. 61103167).